\documentclass[twocolumn,aps,prl,showpacs,raggedbottom,nobalancelastpage,amssymb,superscriptaddress]{revtex4-1}

\usepackage{graphicx}
\usepackage{amsmath}
\usepackage{amssymb}
\usepackage{bm}
\usepackage[usenames]{color}
\usepackage{amsfonts}
\usepackage{appendix}
\usepackage{dsfont}

\newcommand{\nj}[1] { \langle n_{#1} \rangle }
\newcommand{\ave}[1]{\left\langle #1\right\rangle}
\newcommand{\ket}[1]{| #1 \rangle}
\newcommand{\bra}[1]{\langle #1 |}

\def \pj{\langle p_j \rangle}

\begin{document}

\title{Enhanced compressibility due to repulsive interaction in the Harper model}

\author{Yaacov E.~Kraus}
\affiliation{Department of Physics, Bar-Ilan University, Ramat-Gan 52900, Israel}
\affiliation{Department of Condensed Matter Physics, Weizmann Institute of Science, Rehovot 76100, Israel}
\author{Oded Zilberberg}
\affiliation{Institute for Theoretical Physics, ETH Zurich, 8093 Z{\"u}rich, Switzerland}
\author{Richard Berkovits}
\affiliation{Department of Physics, Bar-Ilan University, Ramat-Gan 52900, Israel}

\begin{abstract}
We study the interplay between repulsive interaction and an almost staggered on-site potential in one-dimension. Specifically, we address the Harper model for spinless fermions with nearest-neighbor repulsion, close to half-filling. Using density matrix renormalization group (DMRG), we find that, in contrast to standard behavior, the system becomes more compressible as the repulsive interaction is increased. By generating a low-energy effective model, we unveil the effect of interactions using mean-field analysis: the density of a narrow band around half-filling is anti-correlated with the on-site potential, whereas the density of lower occupied bands follows the potential and strengthens it. As a result, the states around half-filling are squeezed by the background density, their band becomes flatter, and the compressibility increases.
\end{abstract}


\maketitle

There has been much interest in the influence of electron-electron (e-e) interactions on the compressibility of electronic systems. This interest is motivated by the intricate many-body physics revealed by the behavior of the compressibility, as well as by the technological challenge of building field effect transistors with larger capacitance, essential for lower power consumption and quicker clock rates~\cite{li11,tinkl12}.

The compressibility of an electronic system, i.e.~the change in the number of electrons residing in a system as the chemical potential is varied, can be measured via capacitive coupling to another metallic system. Alternatively, the system can be weakly coupled to a plunger gate and leads. Jumps in the current that passes through the leads as a function of the gate voltage count the number of electrons in the system as a function of the chemical potential. In the context of quantum dots, this is known as the addition spectrum~\cite{alhassid00}.

Compressibility is relevant also for other highly controlled many-body systems such as molecular manipulation on metal surfaces~\cite{Manoharan:2012} and ultracold atoms and ions in optical lattices~\cite{BlochReview,Roati08,Bloch:2013,Ketterle:2013}. In optical lattices, transport measurements are challenging. Nevertheless, squeezing the trapping potential acts on the density as a variation of the chemical potential, revealing the bulk compressibility~\cite{Roscilde09}.

Usually, the influence of repulsive interactions on the ground state of fermionic systems is well described by a mean-field theory, which may be reduced to the classical capacitance of the system. Increasing the repulsive interaction then corresponds to reduced capacitance, i.e.~the system becomes less compressible. This is true even beyond the mean-field treatment, as shown for 1D Luttinger liquids with $K<1$~\cite{Giamarchi}, and in 0D quantum dots~\cite{alhassid00}. Interestingly, there are several unexplained counter examples measured in disordered semiconducting dots~\cite{ashoori92,zhitenev97}.

In this work, we present a simple 1D system for which the compressibility \emph{increases} with weak e-e interaction. Specifically, we study the Harper (or Aubry-Andr\'{e}) model~\cite{Harper:1955,AA} of spinless fermions close to half filling with nearest-neighbor repulsive interaction. The on-site potential is spatially modulated with a frequency of an almost two lattice-sites period, corresponding to a fast modulation with a slow envelope. Using DMRG, we numerically extract the inverse compressibility as a function of the density, and find it decreasing with the interaction strength. We analytically show that this effect results from the presence of a flat band at half-filling, which is composed of a superlattice of states that reside at the valleys of the potential envelope. The repulsive interaction from occupied lower bands squeezes these valley states, and accordingly, the central band flattens and its compressibility increases.

The tight-binding Harper model for spinless fermions with nearest-neighbor repulsive interaction is
\begin{align} \label{Eq:H1D}
    H = & \sum_{j = 1}^L \left[ t c_j^\dag c_{j + 1}^{\phantom{\dag}} + h.c + \lambda \cos (2\pi b j + \phi) n_j^{\phantom{\dag}} + U n_j^{\phantom{\dag}} n_{j+1}^{\phantom{\dag}} \right] ,
\end{align}
where $c_j$ is the single-particle annihilation operator at site $j$, $n_j = c^\dag_j c_j$ is the density, $t$ is a real hopping amplitude, $\lambda > 0$ controls the amplitude of the on-site potential, and $U > 0$ is the strength of the repulsive interaction. The potential is cosine modulated in space with frequency $b$, and $\phi$ is an arbitrary phase factor.

The Harper model is a wellspring of physical phenomena, and is therefore under continuous study. For example, when the modulation frequency $b$ is an irrational number, and in the absence of interaction, a metal-insulator transition takes place as a function of the potential strength at $\lambda = 2t$~\cite{AA,hiramoto92,Svetlana,Roati08,Yoav09,Chabe08,Modugno10}. Much effort has gone into understanding the influence of interaction on this transition~\cite{vidal99,vidal01,schuster02,iyer13}. Recently, it was also found that for an irrational $b$, the Harper model is topologically nontrivial, and may have topological boundary states~\cite{us,us2,us3,us5}.

\begin{figure}
\centering
\includegraphics[width=\columnwidth]{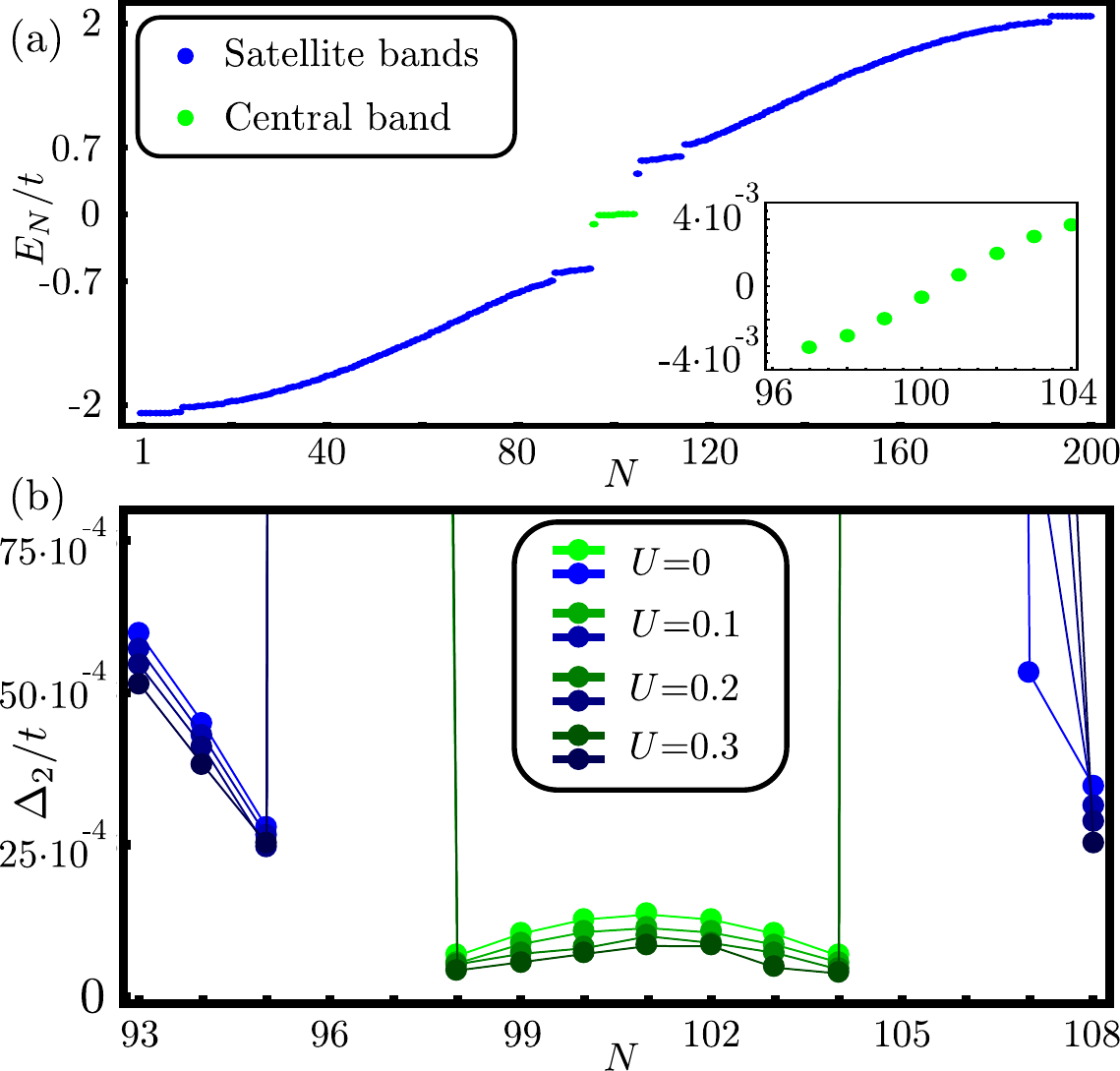}
\caption{\label{Fig1} %
\textit{The effect of interaction}: (a) The single-particle spectrum of the Harper model [cf.~Eq.~\eqref{Eq:H1D} with U=0] with an open boundary condition, for $L=200$, $t=1$, $\lambda=0.7$, $\phi=0.7\pi$, and $b=\sqrt{30}$, which corresponds to $\epsilon\approx -0.023$. The central band (bright green) is almost flat, as seen in the blow-up inset~\cite{note:boundary}. (b) The inverse compressibility as a function of the number of particles [cf.~Eq.~\eqref{Eq:Delta2}], obtained by DMRG. Surprisingly, the central band becomes more compressible as a function of interaction strength, $U$.}
\end{figure}

Here, we are interested in the effect of the interplay between the inhomogeneous potential and the interaction on compressibility. Therefore, in the following, we assume that we are in the metallic phase, i.e.~$\lambda < 2t$. Moreover, we consider the cases of $b \, {\rm mod} \, 1= 1/2+\epsilon$ with $|\epsilon| \ll 1/2$, be it rational or irrational. The striking property of such $b$ is that in the vicinity of half-filling, the energy spectrum is composed of an almost-flat central band separated from the other bands by large gaps, as depicted in Fig.~\ref{Fig1}(a). Therefore, even weak interaction may generate interesting phenomena.

The inverse compressibility of a system with $N$ particles, $\Delta_2(N)$, is defined as the change in the chemical potential due to the insertion of the $N^{\rm th}$ particle. For many-body systems, it is given by
\begin{align} \label{Eq:Delta2}
\Delta_2(N) = \mathcal{E}(N) - 2\mathcal{E}(N-1) + \mathcal{E}(N-2) \,,
\end{align}
where $\mathcal{E}(N)$ is the system's many-body ground-state energy with $N$ particles. For noninteracting systems at zero temperature, $\Delta_2(N; U = 0) = E_N - E_{N-1}$, where $E_N$ is the $N^{\rm th}$ single-particle eigenenergy.

A finite sized Harper model can be thought of as a quasi-disordered 1D quantum (anti-)dot. At low temperatures, the inverse compressibility of a disordered quantum dot is usually described by the CI model, which has been shown to fit experimental measurements very well~\cite{alhassid00}.  According to this model, $\Delta_2(N) = \Delta_2(N;U=0) + e^2/C$, where $C \approx L$ is the total capacitance, and $e^2\approx U$. Thus, an increase in $U$ increases $\Delta_2$.

We extract $\Delta_2(N)$ of our interacting system using DMRG~\cite{white92,white93}. We choose $b=\sqrt{30}$ and $\phi=0.7 \pi$. The former corresponds to $\epsilon \approx -0.023$. The system is of length $L=200$, with $N=91,92,\ldots ,108$ electrons. For $t=1$, the potential amplitude is $\lambda=0.7$, which creates a central band that is very flat, but keeps $\Delta_2$ greater than the numerical accuracy. Interaction strengths of $U=0.1,0.2,0.3$ are considered. The boundary condition is open, since it significantly improves accuracy~\cite{white92,white93}. Keeping $384$ target states, we extract the ground-state energy $\mathcal{E}(N)$ for each $N$. The numerically obtained $\Delta_2(N)$, using Eq.~\eqref{Eq:Delta2}, is depicted in Fig.~\ref{Fig1}(b). The accuracy of $\Delta_2$ drops as $U$ increases, and is about $\pm 3 \cdot 10^{-4}t$ for $U=0.3$. Strikingly, the inverse compressibility decreases with increasing $U$. This implies that the underlying physics is very different than the one of the CI model.

\begin{figure}
\centering
\includegraphics[width=\columnwidth]{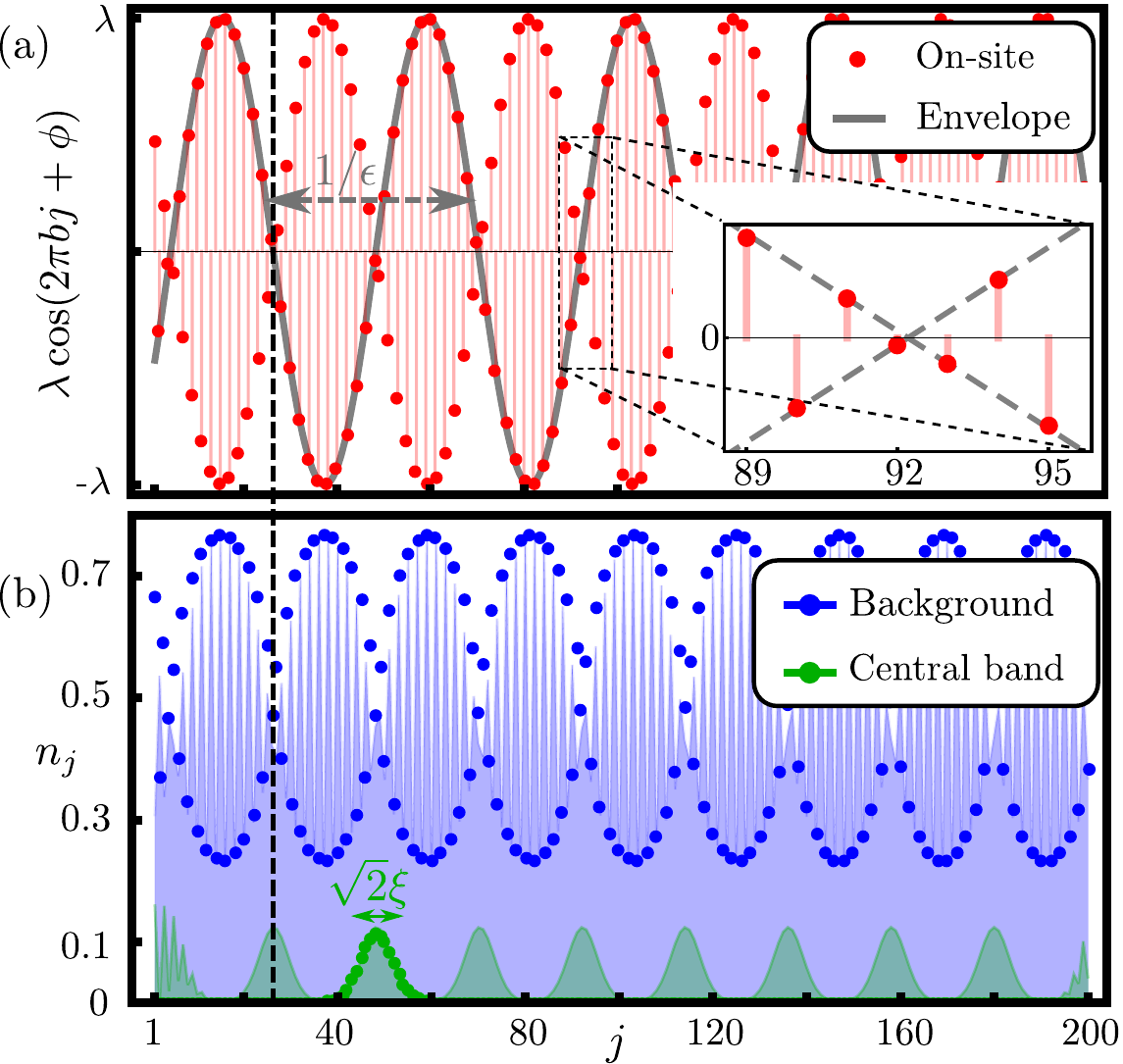}
\caption{\label{Fig2} %
\textit{Modulated potential}: (a) The on-site potential (red dots) is a product of a fast alternating part and a slow envelope (solid gray line), corresponding to $\cos(2\pi \epsilon j + \phi)$. Inset: at its valleys, the potential is linearly approximated. (b) The density of the central band (green), and the background density of the occupied states below it (blue). The filling corresponds to numerical results, whereas the dots correspond to the analytical expressions [cf.~Eqs.~\eqref{Eq:zero_state} and \eqref{bg_density}, respectively]. The central band is composed of waves of hybridized Gaussians that form a superlattice. Remarkably, the states of the central band reside in the potential valleys, whereas the background density follows the potential peaks.}
\end{figure}

Remarkably, we can reproduce this behavior analytically. First we study the noninteracting case, i.e.~$U = 0$. The on-site cosine modulation can be rewritten as $\lambda\cos(2\pi b j + \phi) = \lambda\cos(2\pi \epsilon j + \phi)(-1)^j$. Since $\epsilon \ll 1$, the potential is locally oscillating with modulation frequency $b =1/2$, while being subject to an amplitude envelope, $\lambda(j)$, varying slowly in space with wavelength $1/\epsilon$, see Fig.~\ref{Fig2}(a).

We postulate that the low-energy physics around $E=0$, and in particular that of the central flat band, is governed by states that minimize both the kinetic and potential energies. The potential energy is minimized by states that reside within the valleys of the potential, where the envelope vanishes, i.e.~in the vicinity of $j \approx l_z$, where $2\pi\epsilon l_z +\phi\approx (\mathbb{Z}+1/2)\pi$. Within the $z^{\rm th}$ valley of the potential, we can linearly approximate the envelope, $\cos (2\pi \epsilon j + \phi) \approx 2\pi|\epsilon| (j-l_z)s_z$, where $s_z = -{\rm sign}\left[\sin(2\pi \epsilon l_z+\phi)\right] = \pm 1$. The effective Hamiltonian for a particle confined to the valley is therefore
\begin{eqnarray} \label{Eq:Hvalley}
&& H^{\rm valley} = \sum_{j = 1}^L \left[ t c_j^\dag c_{j + 1}^{\phantom{\dag}} + h.c. + s_z 2\pi|\epsilon|\lambda (-1)^j (j-l_z) c_j^\dag c_j^{\phantom{\dag}} \right] \nonumber \\
&& = \int_0^\pi \frac{dk}{2\pi/L}{\bf \psi}^\dag_k \left[2t\cos(k)\sigma_x + s_z 2\pi|\epsilon| \lambda(\hat{p}_k-l_z) \sigma_z \right]{\bf \psi}_k\,,
\end{eqnarray}
where $\hat{p}_k=i\partial_k$, $\sigma_i$ are Pauli matrices, and ${\bf \psi}_k=(c_{{\rm e}k},c_{{\rm o}k})^T$ is the sublattice psuedospinor that splits the lattice into even and odd sites, according to $c_{{\rm{e}}k}= \sqrt{2/L}\sum_{j = 1}^{L/2}e^{ik2j}c_{2j}$ and $c_{{\rm{o}}k}= \sqrt{2/L}\sum_{j = 1}^{L/2}e^{ik(2j-1)}c_{2j-1}$.

Around zero kinetic energy, we linearize $\cos(k) \approx -(k-\pi/2)$. Now, using the rotation $\hat{T} = (1 + i s_z\sigma_x)/\sqrt{2}$, we rewrite the Hamiltonian in a supersymmetric form
\begin{align} \label{Eq:Hvalley_SS}
H^{\rm valley} = \sqrt{8}\frac{t}{\xi} \int_0^\pi \frac{dk}{2\pi/L}( \hat{T} {\bf \psi}_k )^\dag \left(
\begin{array}{cc} 0 & a_k^\dag \\ a_k & 0 \end{array}
\right)( \hat{T} {\bf \psi}_k )\,,
\end{align}
where $a_k = -(k - \pi/2)\xi/\sqrt{2} + i (\hat{p}_k-l_z)/\sqrt{2}\xi$, and $\xi^2 = t/(\pi\lambda |\epsilon|)$. Since $a_k$ satisfies $[a_k, a_k^\dag] = 1$, it is a ladder operator. Remarkably, this momentum-space Hamiltonian is similar to that of the 2D massless Dirac equation in the presence of a perpendicular magnetic field in Landau gauge. Using the ladder operators, we find that the energy spectrum of $H^{\rm valley}$ is $\pm \sqrt{8n}t/\xi$, where $n = 0, 1, \ldots$~\cite{supmat}. In particular, there is a zero-energy solution with eigenstate
\begin{align} \label{Eq:zero_state}
|l_z\rangle \approx (\pi\xi^2)^{-1/4} \textstyle{\sum_{j=1}^L} (s_z)^j \mathcal{S}_j e^{-(j - l_z)^2/2\xi^2} |j\rangle\,,
\end{align}
where $|j\rangle = c_j^\dag |{\rm vacuum}\rangle$, $\mathcal{S}_j = \sqrt{2} \cos \left(j\pi/2-\pi/4\right)=\ldots,1, 1,-1,-1, 1, 1,\ldots$, and we used the fact that $\xi \gg 1$. This wave function is confined to a Gaussian of width $\xi$ around $l_z$. Notably, the wave functions of the excited states are also confined with the same Gaussian, similar to the eigenstates of the harmonic oscillator~\cite{supmat}.

Turning back to the original noninteracting Hamiltonian, there is a superlattice of valleys, each with its corresponding zero-energy state. We expect these states to hybridize and form the central band. The $|l_z\rangle$ states form a basis for this subspace, since $\langle l_z|l_{z\pm1}\rangle = 0$ and $| \langle l_z|l_{z'}\rangle | \leq e^{-(l_z-l_{z'})^2/2\xi^2} \ll 1$. We can therefore project the Hamiltonian to this subspace. The projected Hamiltonian is given by the matrix elements $\langle l_z|H(U=0)|l_{z'}\rangle$. The diagonal elements $z'=z$ vanish, since $|l_z\rangle$ is of zero energy. The Gaussian decay implies that for $|z'-z|\geq 2$, the matrix elements are negligible. We are therefore left only with $\langle l_z|H(U=0)|l_{z\pm 1}\rangle$, namely, hopping between neighboring valleys. The resulting effective Hamiltonian for the central band is~\cite{supmat}
\begin{align} \label{eff_model}
&H^{\rm central} = -\bar{t} \, \textstyle{\sum_{z=1}^{L_z}} (-1)^z c_{l_z}^\dag c_{l_{z + 1}}^{\phantom \dag}+ h.c.\,,
\end{align}
where $L_z = \lfloor 2|\epsilon| L \rfloor$ is the number of valleys, and $\bar t \approx e^{-\xi^2/(4\xi^2\epsilon)^2}\big(2te^{-1/4\xi^2}\sinh[(4\xi^2|\epsilon|)^{-1}] - \lambda e^{-\pi^2 \epsilon^2 \xi^2}\big)$.

Notably, we obtain $\bar{t} \approx 0.0012$, which is slightly smaller than the numerically observed $\bar t\approx 0.0019$, [cf.~inset of Fig.~\ref{Fig1}(a)]. The small discrepancy arises from using the linear approximation of the potential also between the valleys, leading to a too-fast decay of the wave function. Substituting $1.16\xi^2$ for $\xi^2$ in the expression of $\bar{t}$, corrects the bandwidth.

For a periodic boundary condition, the eigenstates of $H^{\rm central}$ are plane waves $|k\rangle = L_z^{-1/2} \sum_{z=1}^{L_z} S_z e^{ikz} |l_z\rangle$ with spectrum $E^{\rm central}(k) = -2\bar{t}\cos k$, where $k = 2\pi n/L_z$ with $n=1,...,L_z$. Note that these are plane-waves of valley Gaussians, as can be seen from Fig.~\ref{Fig2}(b), which depicts the total density of the central band.
Notably, the bandwidth of the central band, $4\bar{t}$, is much smaller than the gap to the bands of the first excited states $\sqrt{8}t/\xi$, as seen in Fig.~\ref{Fig1}(a). Therefore weak interaction and low temperatures will not mix it with the other bands.

Turning on the repulsive interaction $U$, the effective model of the central band enables us to describe the increase in compressibility using mean-field theory. Here, $\sum_j n_{j+1}n_j$ is approximated by $\sum_j [ \big(\nj{j+1} + \nj{j-1} \big)n_j -\nj{j}\nj{j+1}- \pj c_{j+1}^\dag c_{j}^{\phantom \dag}- \pj^* c_{j}^\dag c_{j+1}^{\phantom \dag}+ |\pj|^2 ]$, with $\nj{j}$ as the background density, and $\pj = \ave{c_{j}^\dag c_{j+1}^{\phantom \dag}}$ as the background exchange energy, both created by the occupied satellite bands below the central band. The constant terms do not contribute to $\Delta_2(N)$, and will therefore be ignored. The mean-field approximation adds effective single-particle on-site potential and hopping, which are modified according to the background density and exchange energy.

We therefore turn to estimate $\nj{j}$ and $\pj$, and begin with solving the simplest Hamiltonian of Eq.~\eqref{Eq:H1D} with $U = \epsilon = 0$. This Hamiltonian describes a uniform staggered potential $(-1)^j\lambda\cos\phi$. Its spectrum is gapped, unless the staggered potential is turned off at $\phi = \pi/2$. If the lower band is fully occupied, then the many-body density is also staggered, $\nj{j}|_{\epsilon = 0} = 1/2 - (-1)^j \bar{n}(\lambda\cos\phi/2t)$, whereas the many-body exchange energy is constant in space $\pj|_{\epsilon = 0}=\bar{p}(\lambda\cos\phi/2t)$, where
\begin{align} \label{Eq:R}
\bar{n}(x) & = \pi^{-1} \textrm{sign}(x) \textrm{K}(-x^{-2})\,,\\
\bar{p}(x)&= -\pi^{-1} |x| \left[\textrm{E}(-x^{-2})- \textrm{K}(-x^{-2})\right]\,,
\label{Eq:Q}
\end{align}
and $\textrm{K}(x)$ and $\textrm{E}(x)$ are the complete elliptical integrals of the first and second kind, respectively~\cite{supmat}.

For $\epsilon \neq 0$, the on-site potential corresponds locally to $(-1)^j$, while $\lambda\cos\phi$ varies slowly in space. Therefore, we expect that the above expressions remain valid locally and vary slowly in space,
\begin{align}
\label{bg_density}
\nj{j} & \approx 1/2 - (-1)^j \bar{n}\left[\lambda\cos(2\pi \epsilon j + \phi)/2t\right]\,,\\
\pj & \approx \bar{p}\left[\lambda\cos(2\pi \epsilon j + \phi)/2t\right]\,.
\label{bg_exchange}
\end{align}
Fig.~\ref{Fig2}(b) depicts the background density $\nj{j}$ obtained both numerically and analytically, according to Eq.~\eqref{bg_density}, and they fit very well. It can be seen that $\nj{j}$ follows the cosine modulation. Therefore, between the valleys, $\nj{j} \approx 1/2 - (-1)^j \bar{n}(\lambda/2t)\cos(2\pi \epsilon j + \phi)$, to first approximation. The background exchange energy, $\pj$, is approximately uniform in space, and thus, $\pj \approx \bar{p}(\lambda/2t)$.

\begin{figure}
\centering
\includegraphics[width=\columnwidth]{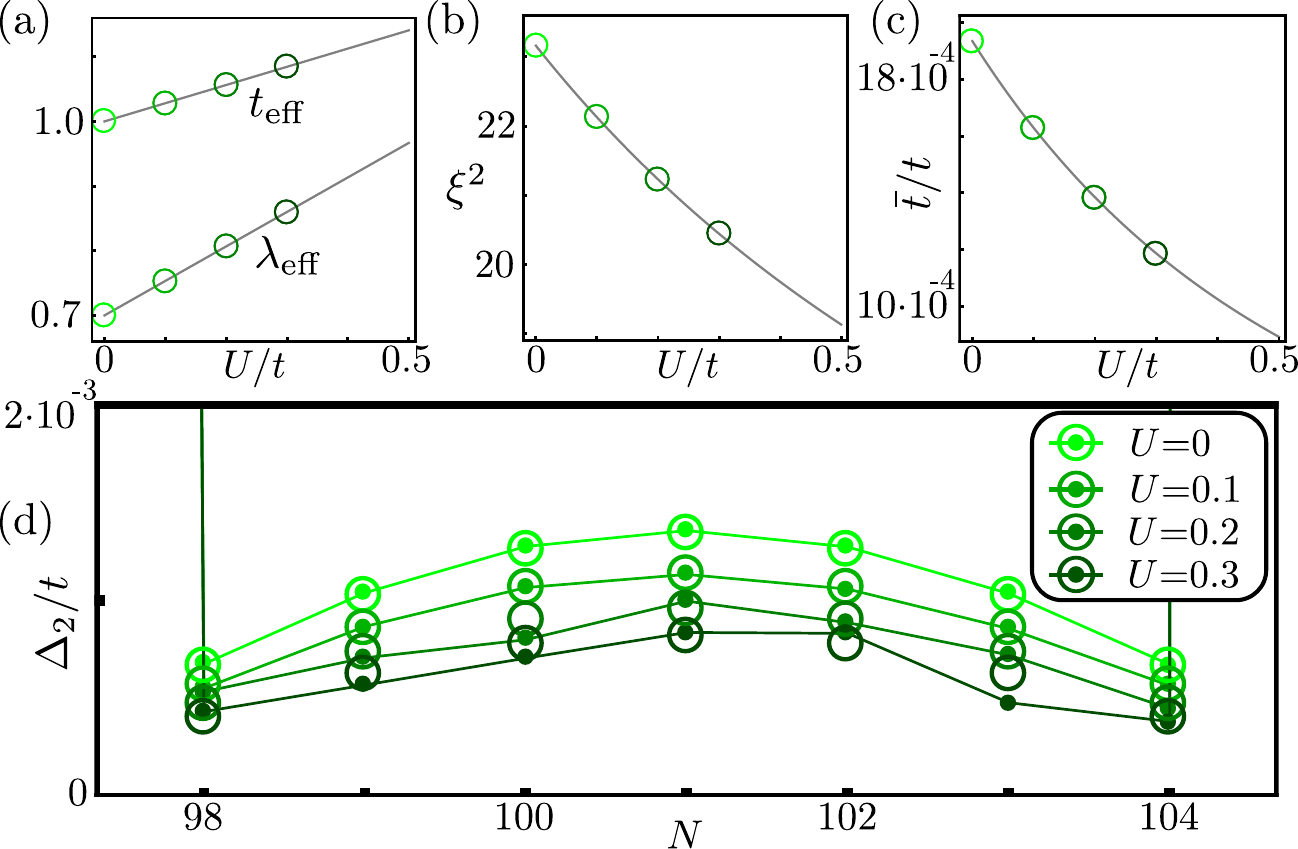}
\caption{\label{Fig3} %
\textit{Mean-field theory:} (a)-(c) The effect of interaction between the central band and the background density, parametrized by the strength $U$, on: (a) the effective on-site and hopping amplitudes, $\lambda_{\rm eff}$ and $t_{\rm eff}$, for bare $t = 1$ and $\lambda = 0.7$ [cf.~Eq.~\eqref{Eq:MF}]; (b) the width of the single-valley Gaussian, $\xi^2$ [cf.~Eq.~\eqref{Eq:zero_state}]; and (c) the resulting hopping amplitude $\bar{t}$ of the central band's effective model [cf.~Eq.~\eqref{eff_model}]. (d) The inverse compressibility of the central band obtained by the effective model of the central band (open circles) compared to that obtained by DMRG (solid lines and dots, cf.~Fig.~\ref{Fig1}). Increasing interaction corresponds to darker (green) shades.  }
\end{figure}

Substituting these simplifications in the mean-field approximation of $H$, we find that the background density increases the modulated on-site potential, and the exchange energy enhances the hopping,
\begin{align} \label{Eq:MF}
H^{\rm MF} = & \sum_{j = 1}^L \left[ t_{\rm eff} c_j^\dag c_{j + 1} + h.c + \lambda_{\rm eff}\cos (2\pi b j + \phi) n_j + U n_j \right]\, ,
\end{align}
where $\lambda_{\rm eff} = \lambda + 2U\bar{n}(\lambda/2t)$ and $t_{\rm eff} = t + U\bar{p}(\lambda/2t)$. Like $H(U=0)$, $H^{\rm MF}$ has a central band of superlattice states. Nevertheless, the width of the valley states, $\xi$, and their hopping amplitude, $\bar{t}$, are here determined by $\lambda_{\rm eff}$ and $t_{\rm eff}$, rather than by $\lambda$ and $t$. Although both $\lambda_{\rm eff}$ and $t_{\rm eff}$ increase with $U$, $\lambda_{\rm eff}$ grows faster. Therefore, $\xi = \xi(t_{\rm eff}/\lambda_{\rm eff})$ decreases as a function of $U$, making the Gaussians squeezed. Consequently, $\bar{t}$ also reduces, and the central band becomes narrower, see Figs.~\ref{Fig3}(a)-(c). Intuitively, it is caused by the fact that the background density follows the on-site potential, whereas the states of the central bands are localized in its valleys. Therefore, the repulsion from background density squeezes the Gaussians and reduces their overlap.

In order to recover the enhanced compressibility in the interacting case, one can diagonalize the effective noninteracting model of the central band $H^{\rm central}$ [cf.~Eq.~\eqref{eff_model}] with $\bar{t}(t_{\rm eff},\lambda_{\rm eff})$, and extract $\Delta_2$. Fig.~\ref{Fig3}(d) depicts $\Delta_2(N,U)$ that is obtained following this procedure. Clearly, it fits nicely to the one observed by DMRG. We note that the low-energy excitations of the satellite bands are also composed of valley states, and therefore they also have decreasing compressibility, as implied by Fig.~\ref{Fig1}(b). However, since the gap that separates them from the background states is much smaller, they will mix for much weaker interaction and lower temperatures.

To summarize, we studied the Harper model of spinless fermions at half-filling with nearest-neighbors repulsive interaction, for modulation frequency $b = 1/2 + \epsilon$. We find that in the absence of interaction, a narrow band appears at zero energy, which is separated by a gap from the other bands. This band is composed of states that are localized within the superlattice of valleys of the potential. Conversely, the lower occupied bands form a background density that follows the on-site potential. The strength of the on-site potential that is experienced by the valley states is increased by the repulsion from the background, $\lambda_{\rm eff} > \lambda$. As a result, these states become narrower, i.e., their overlap diminishes, and their inverse compressibility $\Delta_2$ decreases. This unexpected effect may hint for an explanation for the above mentioned experiments~\cite{ashoori92,zhitenev97}. Furthermore, it provides new insight on the interplay between localization and e-e interaction. Last, our model is readily implemented in existing technology of cold atoms in optical lattices~\cite{BlochReview,Roati08}. Notably, using hard-core bosons, rather than spinless fermions, will increase the effect, since now the effective hopping is reduced by interaction $t_{\rm eff} < t$.

\begin{acknowledgments}
We would like to thank E.~Berg, S.~Huber, and R.~Chitra for useful discussions.
Financial support from the Israel Science Foundation (Grant 686/10), the US-Israel Binational Science Foundation, the Minerva foundation and the Swiss National Foundation is gratefully acknowledged.
\end{acknowledgments}


\begin{thebibliography}{31}
\bibitem[{not()}]{note:boundary}
\bibinfo{note}{The subgap states that appear near the bands are topological
  boundary states~\cite{us}.}
\end{thebibliography}

\newpage
\cleardoublepage
\newpage
\begin{center}
\textbf{\large SUPPLEMENTAL MATERIAL}
\end{center}
\vspace{5mm}

\setcounter{enumi}{1}
\setcounter{equation}{0}
\renewcommand{\theequation}{\Roman{enumi}.\arabic{equation}}

\section{I. Spectrum of valley Hamiltonian}

In the main text, we presented the approximated Hamiltonian of the noninteracting case within a valley of the potential, and its Fourier transform, $H^{\rm valley}$ [cf.~Eq.~(3) of the main text]. By rotating it in the sublattice space with the rotation operator $\hat{T} = (1 + i s_z\sigma_x)/\sqrt{2}$, and introducing the ladder operator $a_k = -(k - \pi/2)\xi/\sqrt{2} + i (\hat{p}_k-l_z)/\sqrt{2}\xi$, we turned $H^{\rm valley}$ into supersymmetric form [cf.~Eq.~(4) of the main text]. Here we derive its energy spectrum and eigenstates.

A supersymmetric Hamiltonian has a zero-energy eigenstate. We find it by solving the differential equation $a_k \varphi_0(k) = 0$, which gives \begin{align} \label{Eq:phi_0}
\varphi_0(k) = \sqrt{L_k} (\xi^2/\pi)^{1/4} e^{-il_z k}e^{- (k-\pi/2)^2 \xi^2/2} \,,
\end{align}
where $L_k = 2\pi/L$. For higher energy eigenstates, we define the functions $\varphi_n(k) = (a_k^\dag)^n \varphi_0(k) /\sqrt{n!}$, which satisfy the relations $a_k \varphi_n(k) = \sqrt{n} \varphi_{n-1}(k)$ and $a_k^\dag \varphi_n(k) = \sqrt{n+1} \varphi_{n+1}(k)$. Basically, these functions are the eigenstates of the harmonic oscillator, shifted by $\pi/2$ and multiplied by the phase factor $e^{il_z l}$.

The corresponding eigenstates of $H^{\rm valley}$ are
\begin{align} \label{Eq:npm}
\bar{\varphi}_{n,\pm}^\dag &= \int_0^\pi \frac{dk}{\sqrt{2}L_k} ( \hat{T} {\bf \psi}_k )^\dag
    \left( \begin{array}{c}
         \varphi_n(k) \\
         \pm\varphi_{n-1}(k) \end{array} \right) \nonumber \\
&= \int_0^\pi \frac{dk}{2L_k} {\bf \psi}_k^\dag
    \left( \begin{array}{c}
         \varphi_n(k) \pm is_z \varphi_{n-1}(k) \\
         is_z \varphi_n(k) \pm \varphi_{n-1}(k) \end{array} \right) \,.
\end{align}
Note that for $n=0$ there is only one solution, with $\varphi_{-1}(k) \equiv 0$, and we multiply this expression by $\sqrt{2}$. Their energies are $E^{\rm valley}_{n,\pm} = \pm 2t\sqrt{2n}/\xi$, as mentioned in the main text. Notably, in its supersymmetric form, $H^{\rm valley}$ anticommutes with $\sigma_z$, and thus, the states come as particle-hole pairs, $\bar{\varphi}_{n,-}^\dag = \sigma_z \bar{\varphi}_{n,+}^\dag$. The only unpaired state is the zero-energy state, $\bar{\varphi}_0$, which is also protected at zero energy, as long as the anticommutation holds.

We can see that all the eigenstates have the same phase factor $e^{-il_z k}$ and are confined by the same Gaussian $e^{- (k-\pi/2)^2 \xi^2/2}$. Therefore, in real space, they are also confined by a Gaussian $e^{-(j - l_z)^2 /2\xi^2}$ around the node $\l_z$, and are accompanied by the phase factor $e^{ij\pi/2} = i^j$.

We are mostly interested in the wave function of the zero-energy state,
\begin{align} \label{Eq:lz}
\bar{\varphi}_0^\dag &= \int_0^\pi \frac{dk}{\sqrt{2}L_k}
\begin{array}{c} (c_{{\rm e}k}^\dag,c_{{\rm o}k}^\dag) \\ \phantom{\varphi} \end{array}
    \left( \begin{array}{c}
         \varphi_0(k) \\
         is_z \varphi_0(k) \end{array} \right) \\
&= \sum_{j=1}^{L/2} \int_0^\pi \frac{dk}{\sqrt{2\pi L_k}} [ \varphi_0(k) e^{ik2j} c_{2j}^\dag + is_z \varphi_0(k) e^{ik(2j-1)} c_{2j-1}^\dag] \nonumber \\
&= (\pi\xi^2)^{-1/4} e^{-il_z\pi/2} \textstyle{\sum_{j=1}^{L/2} (-1)^j} \nonumber \\
& \qquad \times \left( e^{-(2j-l_z)^2/(2\xi^2)} c_{2j}^\dag - s_z e^{-(2j-1 - l_z)^2/(2\xi^2)} c_{2j-1}^\dag \right) \nonumber \,.
\end{align}
By defining $\ket{l_z} \equiv \bar{\varphi}_0^\dag \ket{\textrm{vacuum}}$ and $\mathcal{S}_j = \sqrt{2} \cos \left(j\pi/2-\pi/4\right)=\ldots,-1, -1,1,1, -1, -1,\ldots$, and omitting the global phase factor, we obtain Eq.~(5) of the main text.


\section{II. Hopping term of central-band Hamiltonian}

The central band of the noninteracting spectrum is our main interest. This band is composed of multiple single-valley states $\ket{l_z}$ [cf.~Eq.~(5) of the main text], which overlap and thus hop. In order to get an effective Hamiltonian for this band, we project the full noninteracting Hamiltonian $H(U = 0)$ [cf.~Eq.~(1) of the main text] onto the space of the $\ket{l_z}$ states. In the main text we have seen that the only relevant matrix elements of the projection are $\bra{l_z} H(U=0) \ket{l_{z\pm1}}$, and here we evaluate them.

Recall that $l_{z\pm1} = l_z \pm 1/2|\epsilon|$, and accordingly we can substitute $\cos(2\pi\epsilon j + \phi) = \pm s_z \cos[ 2\pi\epsilon(j - l_z \mp 1/4|\epsilon|)]$. We can now evaluate
\begin{eqnarray} \label{Eq:lzHlz1}
&& \bra{l_z} H(U=0) \ket{l_{z\pm1}} = \textstyle{ \sum_{j = 1}^L } \bra{l_z}t c_j^\dag c_{j + 1}^{\phantom{\dag}} + h.c\ket{l_{z\pm1}} \nonumber \\
&& \qquad \pm s_z \lambda \cos[ 2\pi\epsilon(j - l_z \mp 1/4\epsilon)] \bra{l_z} c_j^\dag c_j \ket{l_{z\pm1}} \nonumber \\
&& = \frac{s_z}{\sqrt{\pi}\xi} \sum_{j=1}^L \left[ \pm \frac{\lambda}{2} e^{-[(n-l_z)^2 + (n-l_z \mp 1/2|\epsilon|)^2]/(2\xi^2)} \nonumber \right. \\
&& \qquad \qquad \times ( e^{i2\pi\epsilon (n - l_z \mp 1/4|\epsilon|)} + e^{-i2\pi\epsilon (n - l_z \mp 1/4|\epsilon|)} ) \nonumber \\
&& \qquad + t e^{-[(n-l_z)^2 + (n - 1 -l_z \mp 1/2|\epsilon|)^2]/(2\xi^2)} \nonumber \\
&& \qquad \left. - t  e^{-[(n-l_z)^2 + (n + 1 -l_z \mp 1/2|\epsilon|)^2]/(2\xi^2)} \right]
\end{eqnarray}
where we used the relation $\mathcal{S}_{j\pm1} = \pm (-1)^j \mathcal{S}_j$. Using also the fact that $\xi \gg 1$, we shift the summation, and obtain
\begin{eqnarray} \label{Eq:lzHlz2}
&& \bra{l_z} H(U=0) \ket{l_{z\pm1}} \approx \pm s_z e^{-1/(4\xi\epsilon)^2} \times \\
&& \left[ \frac{\lambda}{2} e^{-(\pi\epsilon\xi)^2} \cdot \frac{1}{\sqrt{\pi}\xi} \sum_{j=1}^L \big( e^{-(j - i\pi\epsilon\xi)^2/\xi^2} + e^{-(j + i\pi\epsilon\xi^2)^2/\xi^2} \big) \right. \nonumber \\
&& \left. + te^{-1/(4\xi^2)} \big( e^{-1/(4\xi^2|\epsilon|)} - e^{1/(4\xi^2|\epsilon|)} \big) \cdot \frac{1}{\sqrt{\pi}\xi} \sum_{j=1}^L e^{-j^2/\xi^2} \right] \,. \nonumber
\end{eqnarray}
Performing the Gaussian sums, we obtain Eq.~(6) of the main text, with the corresponding $\bar{t}$.


\section{III. Solving $H$ for $b = 1/2$ and $U=0$}

In the mean-field theory that we use in the main text, we present the many-body density, $\nj{j}|_{\epsilon = 0}$, and exchange energy, $\pj|_{\epsilon = 0}$, of the lower band of the simple noninteracting Hamiltonian with $b=1/2$, i.e.~$U = \epsilon = 0$. Here we derive the expression for $\nj{j}|_{\epsilon = 0}$ and $\pj|_{\epsilon = 0}$.

For $b=1/2$, our Hamiltonian $H$ [cf.~Eq.~(1) of the main text] becomes
\begin{align} \label{Eq:b_half}
H^{\epsilon = 0} &= \textstyle{ \sum_{j = 1}^L } \left[ t c_j^\dag c_{j + 1}^{\phantom{\dag}} + h.c + \lambda \cos\phi (-1)^j n_j \right] \nonumber \\
&= \int_0^\pi \frac{dk}{L_k} {\bf \psi}^\dag_k [ 2t\cos k\,\sigma_x + \lambda\cos\phi\,\sigma_z]{\bf \psi}_k\,,
\end{align}
where ${\bf \psi}_k=(c_{{\rm e}k},c_{{\rm o}k})^T$ is the sublattice pseudospinor, the same as in the main text [cf.~Eq.(3)]. The energy spectrum of $H^{\epsilon = 0}$ is composed of two bands, $E^{\epsilon=0}_{k,\pm} = \pm \sqrt{ 4t^2\cos^2 k + \lambda^2\cos^2\phi }$. The corresponding eigenstates are
\begin{align} \label{Eq:chi_k}
\chi^\dag_{k,\pm} &= \sqrt{L/2} \begin{array}{c} (c_{{\rm e}k}^\dag,c_{{\rm o}k}^\dag) \\ \phantom{c} \end{array}
\left( \begin{array}{c} \chi_{{\rm e}k,\pm} \\ \chi_{{\rm o}k,\pm} \end{array} \right) \\
&= \sum_{j=1}^L \big( \chi_{{\rm e}k,\pm}e^{ik2j} c_{2j}^\dag + \chi_{{\rm o}k,\pm}e^{ik(2j-1)}c_{2j-1}^\dag \big) \,,
\end{align}
where
\begin{align}
\left( \begin{array}{c} \chi_{{\rm e}k,\pm} \\ \chi_{{\rm o}k,\pm} \end{array} \right) = \frac{1}{\sqrt{2E^{\epsilon=0}_{k,\pm}(E^{\epsilon=0}_{k,\pm} - \lambda\cos\phi)}}
\left( \begin{array}{c} 2t\cos k \\ E^{\epsilon=0}_{k,\pm} - \lambda\cos\phi \end{array} \right) \nonumber
\end{align}

The density of an eigenstate is
\begin{eqnarray} \label{Eq:nk}
&& \bra{\chi_{k,\pm}} n_j \ket{\chi_{k,\pm}} = \frac{ \chi_{{\rm e}k,\pm}^{\phantom{ek+}2} + \chi_{{\rm o}k,\pm}^{\phantom{ek+}2} }{2} + (-1)^j \frac{ \chi_{{\rm e}k,\pm}^{\phantom{ek+}2} - \chi_{{\rm o}k,\pm}^{\phantom{ek+}2} }{2} \nonumber \\
&& \qquad = \frac{1}{2} \mp \frac{1}{2} \frac{ {\rm sign}(\lambda\cos\phi) } {\sqrt{1 + (\lambda\cos\phi/2t)^{-2}\cos^2 k}} \,,
\end{eqnarray}
and its exchange energy is
\begin{eqnarray} \label{Eq:pk}
&& \bra{\chi_{k,\pm}} c^\dag_{j+1}c_j \ket{\chi_{k,\pm}} = e^{-ik} \chi_{{\rm e}k,\pm}\chi_{{\rm o}k,\pm} \\
&& \qquad \qquad = \pm \frac{2t}{|\lambda\cos\phi|} \frac{ e^{-ik}\cos k } {\sqrt{1 + (\lambda\cos\phi/2t)^{-2}\cos^2 k}} \,. \nonumber
\end{eqnarray}
If the lower band is fully occupied, then the many-body density is given by
\begin{align} \label{Eq:nj}
\nj{j}|_{\epsilon = 0} &= \int_{-\pi/2}^{\pi/2} \frac{dk}{\pi} \bra{\chi_{k,-}} n_j \ket{\chi_{k,-}} \nonumber \\
&= 1/2 - (-1)^j \bar{n}(\lambda\cos\phi/2t) \,,
\end{align}
where
\begin{align} \label{Eq:nbar}
\bar{n}(x) &= \int_{-\pi/2}^{\pi/2} \frac{dk}{2\pi} \frac{ {\rm sign}(x) }{\sqrt{1 + (1 - \sin^2 k)/x}} \nonumber \\
&= \frac{x}{\pi\sqrt{1 + x^2}} \int_{0}^{\pi/2} \frac{dk}{\sqrt{1 - (1 + x^2)^{-1}\sin^2 k}} \nonumber \\
&= \frac{x}{\pi\sqrt{1 + x^2}} \textrm{K}\left(\frac{1}{1+x^2}\right) \,,
\end{align}
with ${\rm K}(x)$ as the complete elliptical integral of the first kind. If one uses a definition of ${\rm K}(x)$ in which $x$ is not limited to the interval $(0,1)$, then $\bar{n}(x)$ can be further simplified to yield Eq.~(7) of the main text. Similarly, the many-body exchange energy is
\begin{align} \label{Eq:pj}
\pj|_{\epsilon = 0} &= \int_{-\pi/2}^{\pi/2} \frac{dk}{\pi} \bra{\chi_{k,-}} c^\dag_{j+1}c_j \ket{\chi_{k,-}} \nonumber \\
&= \bar{p}(\lambda\cos\phi/2t) \,,
\end{align}
with
\begin{align} 
\bar{p}(x) &= \frac{-1}{2\pi\sqrt{1 + x^2}} \int_{-\pi/2}^{\pi/2} dk \frac{\cos^2 k - i\sin k\cos k}{\sqrt{1 - (1 + x^2)^{-1}\sin^2 k}} \nonumber
\end{align}
The imaginary part of the integral vanishes due its antisymmetry in the interval. Therefore,
\begin{align} \label{Eq:nbar2}
\bar{p}(x) &= \frac{1}{\pi\sqrt{1 + x^2}} \int_0^{\pi/2} dk \left[ \frac{x^2}{\sqrt{1 - (1 + x^2)^{-1}\sin^2 k}} \right. \nonumber \\
& \quad \left. - (1+x^2)\sqrt{1 - (1 + x^2)^{-1}\sin^2 k} \right] \\
&= \frac{x^2}{\pi\sqrt{1 + x^2}} \textrm{K}\left(\frac{1}{1+x^2}\right) - \frac{\sqrt{1 + x^2}}{\pi}{\rm E}\left(\frac{1}{1+x^2}\right) \,. \nonumber
\end{align}
Here ${\rm E}(x)$ is the complete elliptical integrals of the second kind. Again, for unlimited $x$, $\bar{p}(x)$ can be further simplified to yield Eq.~(8) of the main text.

\end{document}